\documentstyle[12pt]{article}
\begin{document}
\begin{center}
{\Large{\bf A Modified Scheme of Triplectic Quantization}}\\

\vspace{0.5cm}
B. Geyer\\
{\it Naturwissenschaftlich-Theoretisches Zentrum und \\
Institut f\"ur Theoretische Physik, Universit\"at Leipzig,\\
Augustusplatz 10/11, 04109 Leipzig, Germany}\\
\vspace{0.5cm}

D.M. Gitman\\
{\it Instituto de Fisica, Universidade de S\~{a}o Paulo\\
Caixa Postal 66318-CEP, 05315-970-S\~{a}o Paulo, S.P., Brazil}\\

\vspace{0.5cm}

P.M. Lavrov\\
{\it Tomsk State Pedagogical University, 634041 Tomsk, Russia}
\end{center}
\vspace{0.5cm}
\centerline{\bf Abstract}
A modified version of triplectic quantization, first introduced by
Batalin and Marnelius, is proposed which makes use of
two independent master equations, one for the action
and one for the gauge functional such that the
initial classical action also obeys that master equation.

\vspace{1cm}
\section{Introduction}
\vspace{1cm}

Lagrangian quantization (LQ) remains one of the most attractive
approaches to quantize gauge theories. Its main advantage is a
direct construction of the quantum effective action avoiding the
long detour of the Hamiltonian approach through canonical
quantization with subsequent integration over momenta in the path
integral, and producing directly the vacuum
expectation value of the S-matrix in the presence of external
sources (see, for example, \cite{GT}). The main difficulty
faced by the implementation of the LQ program lies in the complicated
structure of the
gauge symmetries corresponding to the initial classical action of a theory,
as well as in the ambiguities existing in the choice of possible gauges. It
is well known that gauge theories with closed algebra of
linearly independent generators of gauge transformations -- when considered
in the class of certain admissible (although restricted) gauges -- permit
the quantum effective action $S_{eff}$ to be constructed directly by means
of the Faddeev-Popov (FP) rules \cite{1a}. The overwhelming majority of
physically interesting theories -- such as Yang-Mills theories, theory of
gravity, etc -- belong to the realm of the application of this
quantization method. According to the FP rules, the action $S_{eff}$
depends on an enlarged set of classical fields $\phi$, composed, apart
from the basic classical fields $A$, also by the sets of auxiliary fields
$B$ (Nakanishi-Lautrup fields) related to the gauge-fixing, as well as by
the sets of FP ghost and anti-ghost fields $C,\;\bar{C}$: altogether
they are given by $\phi=(A,B,C,\bar{C})$.
The action $S_{eff}$ constructed according to the
FP rules is invariant under the (global) nilpotent BRST
transformations \cite{2a,3a}. Assigning to each field $\phi$ the
corresponding BRST--source, or anti--field, $\phi^*$, one
observes that the BRST
symmetry results in the nonlinear Zinn-Justin equation  \cite{4a}
determining the quantum action
$S$ which ensures complete description of $S_{eff}$.

Attempts to go beyond the scope of the above mentioned
restrictions on the structure
of gauge symmetries of the initial classical action -- as well as beyond the
restrictions on the above mentioned classes of gauge conditions -- have
resulted in a number of new schemes of LQ, based of the
so-called ${\em master \;equations}$ (analogous to the Zinn-Justin
equation).
Their solutions determine the quantum action $S$,
which satisfies a set of
natural boundary conditions related to the initial classical action. One
of the first schemes of this kind was the Batalin-Vilkovisky (BV) method
 \cite{5a,6a}, which allows to consider gauge theories with
open gauge algebra. In the case of irreducible theories
\cite{5a}, corresponding to a linearly independent set of generators, the
configuration space $\phi$ of the BV method coincides with that of the FP
approach, whereas in the case of reducible theories \cite{6a} one has to
introduce additional (sometimes infinite) sets of auxiliary and ghost
fields. Afterwards the so-called $Sp(2)$ symmetric scheme has been
developed \cite{7a,8a,9a}, in which the ghost and anti--ghost
fields are treated on
equal footing (in contrast to the BV method, where the anti-ghost fields
appear only at the stage of gauge--fixing), and in which the
(in)finite towers
of auxiliary and ghost fields are classified in terms of irreducible
representations of the $Sp(2)$ group. In fact, such a quantization scheme
realizes a combined BRST -- anti-BRST symmetry principle (the latter
symmetry was first discovered in \cite{10a,11a} for Yang-Mills 
theories in the framework of the FP method). Within the $Sp(2)$ scheme, the
quantum action $S$ depends on an extended set of field variables, which
includes, apart from the variables $\phi$ 
also three 
sets of anti--fields, namely, $Sp(2)$--doublets $\phi^*_a,\;a=1,2$ (sources
of BRST and anti-BRST transforms) and $Sp(2)$--scalars $\bar{\phi}$
(sources of mixed transforms).

In a recent work \cite{12a} Batalin and Marnelius presented a
new possible generalization of the $Sp(2)$ method -- related to an additional
extension of the configuration space of the quantum action $S$ -- and aimed
at providing equal treatment for all anti--fields of the $Sp(2)$ approach.
In particular, it is suggested to consider the fields $\pi^a$ (which
parametrize gauges in the path integral of the $Sp(2)$ method) as
anticanonically conjugated to the anti--fields $\bar{\phi}$, with the
corresponding redefinition of the extended anti--brackets. Thus, along
with the $Sp(2)$--singlet $\bar \phi$, the theory also contains an
$Sp(2)$--doublet
$\pi^a$, which thus accounts for the fact that the corresponding scheme
is referred to as triplectic quantization (TQ). 
An essential original point
of the TQ scheme consists in dividing the entire task of constructing the
quantum effective action $S_{eff}$ into the following two steps: first,
the construction of the quantum action $S$, and second the construction
of the corresponding gauge-fixing functional.
Either problem is solved by means of an appropriate master equation.

Despite considering these new ideas as very promising,
as to their concrete realization we propose a different,
modified scheme of the TQ, which -- especially from some
geometrical viewpoint --
 changes the meaning of the latter. Namely, remaining in the same
configuration space of fields, and accepting the idea of a separate treatment
of the two above mentioned actions, we propose to change both systems of
master equations by using a new set of two $Sp(2)$--doublets of generating
operators: $V^a$ and $U^a$. Such a modification is inspired by our experience
in the superfield formulation of the $Sp(2)$ method \cite{13a}, in which the
above operators acquire the geometrical interpretation of the generators of
(super)transformations in a superspace spanned by fields and anti--fields.
In
this approach, the first master equation, determining the quantum action $S$,
is defined by means of the operators $V^a$, whereas the other master equation,
determining the gauge fixing functional $X$,
is defined by means of the operators
$U^a$. As in the original TQ scheme, we may expect that the generating
functional of Green's functions does not depend on the choice of gauge. It is
important to emphasize that within the modified TQ scheme the entire
information contained in the initial classical action of the theory is
conveyed to the quantum effective action via the corresponding boundary
conditions. At the same time, the classical action obeys the first modified
master equation in complete analogy with all previously known schemes of LQ.
The original TQ scheme gives no explicit relation to the initial classical
action. If one assumes that such a classical action occurs, as usual,
in the boundary condition to the solution of the master equation (with
vanishing auxiliary fields and quantum corrections), then this classical
action does not obey the master equation.

The purpose of this paper is to elaborate a complete description of TQ
scheme -- within the framework of the above mentioned
modifications -- re\-ve\-al\-ing, at the same time, the points which make
such a description similar to, or different from, the original TQ scheme.

\section{Main Definitions}

Let us denote by $\phi^A$, $\varepsilon(\phi^A)\equiv\varepsilon_A$,
the complete set of fields which span the configuration space
corresponding to a certain initial gauge theory. The explicit structure
of the fields $\phi^A$ is not essential for the purposes of the following
treatment (for details, see the original papers \cite{5a,6a}).
According to refs.~\cite{7a,12a} we further introduce the set of antifields
$\phi^*_{Aa}$, $\varepsilon(\phi^*_{Aa})=\varepsilon_A+1$; $\bar{\phi}_A$,
$\varepsilon(\bar{\phi}_A)=\varepsilon_A$, accompanied \cite{12a} by the set
of fields $\pi^{Aa}$, $\varepsilon(\pi^{Aa})=\varepsilon_A+1$. In the entire
space of variables $\phi^A$, $\phi^*_{Aa}$, $\pi^{Aa}$, $\bar{\phi}_A$
we define extended antibrackets, given, for any two functionals $F$, $G$,
by the rule
\begin{eqnarray}
 (F,G)^a=\frac{\delta F}{\delta\phi^A}\frac{\delta G}{\delta\phi^{*}_{Aa}}
 +\varepsilon^{ab}
 \frac{\delta F}{\delta\pi^{Ab}}\frac{\delta G}{\delta\bar{\phi}_{A}}
 -(F\leftrightarrow G)(-1)^{(\varepsilon(F)+1)(\varepsilon(G)+1)}\;,
\end{eqnarray}
where $\varepsilon^{ab}$ is the constant antisymmetric second-rank tensor,
$\varepsilon^{12}=1$. Eq.~(1) coincides with the definition of the
extended antibrackets given in the method of triplectic quantization
\cite{12a}, as well as with the definition of the extended antibrackets given
in the method of superfield quantization \cite{13a}. Speaking of the
algebraic properties of the extended antibrackets, we will only mention the
generalized Jacobi identities
\begin{eqnarray}
 ((F,G)^{\{a},H)^{b\}}(-1)^{(\varepsilon(F)+1)(\varepsilon(H)+1)}
 +{\rm cycle}\,(FGH)\equiv 0.
\end{eqnarray}
Here and elsewhere the curly brackets stand for symmetrization over
$a$, $b$.

In solving the functional equations determining the effective action we
make use of the operators $\Delta^a$, $V^a$ and $U^a$
\begin{eqnarray}
 \Delta^a&=&(-1)^{\varepsilon_A}\frac{\delta_l}{\delta\phi^A}
 \frac{\delta}{\delta\phi^{*}_{Aa}}
 +(-1)^{\varepsilon_A+1}\varepsilon^{ab}\frac{\delta_l}{\delta\pi^{Ab}}
 \frac{\delta}{\delta\bar{\phi}_A},\\
 V^a&=&\varepsilon^{ab}\phi^{*}_{Ab}\frac{\delta}{\delta\bar\phi_A}\;,\\
 U^a&=&(-1)^{\varepsilon_A+1}\pi^{Aa}\frac{\delta_l}{\delta\phi^A}\;.
\end{eqnarray}

Notice that the operators $\Delta^a$ have already appeared both within the
scheme of triplectic quantization \cite{12a} and, virtually, within the
scheme of superfield quantization \cite{13a}. Even though the operators
$V^a$ in eq.~(4) differ from the corresponding operators of the triplectic
quantization \cite{12a}, they coincide, at the same time, with the operators
applied in the framework of the $Sp(2)$ method \cite{7a}. The use of the
operators $U^a$ in eq.~(5) (an analog of these operators has been introduced
in the method of superfield quantization) exhibits an essentially new
feature as compared to both the $Sp(2)$ method and the TQ \cite{12a}.

One easily establishes the following algebra of the operators (3)--(5):
\begin{eqnarray}
 &&\Delta^{\{a}\Delta^{b\}}=0,\\
 &&V^{\{a}V^{b\}}=0,\\
 &&\Delta^{\{a}V^{b\}}+V^{\{a}\Delta^{b\}}=0,\\
 &&U^{\{a}U^{b\}}=0,\\
 &&\Delta^{\{a}U^{b\}}+U^{\{a}\Delta^{b\}}=0,\\
 &&V^aU^b+U^bV^a=0,\\
 &&\Delta^aV^b+V^b\Delta^a+\Delta^aU^b+U^b\Delta^a=0.
\end{eqnarray}
The action of the operators $\Delta^a$ (3) on the product of any two
functionals $F$, $G$,
\begin{eqnarray}
 \Delta^a(F\cdot G)=(\Delta^aF)\cdot G+F\cdot(\Delta^a G)
 (-1)^{\varepsilon(F)}
 +(F,G)^a(-1)^{\varepsilon(F)},
\end{eqnarray}
may serve as an independent definition of the extended antibrackets (1).
The action of each of the operators $\Delta^a$, $V^a$ and $U^a$ (3)--(5)
on the extended antibrackets is given by the rule
($D^a=(\Delta^a,V^a,U^a)$)
\begin{eqnarray}
 D^{\{a}(F,G)^{b\}}=(D^{\{a}F,G)^{b\}}-(F,D^{\{a}G)^{b\}}
 (-1)^{\varepsilon(F)}.
\end{eqnarray}

Apart from $\Delta^a$, $V^a$, we also introduce the operators
\begin{eqnarray}
 \bar{\Delta}^a&\equiv&\Delta^a+\frac{i}{\hbar}V^a,\\
 \tilde{\Delta}^a&\equiv&\Delta^a-\frac{i}{\hbar}U^a.
\end{eqnarray}
From eqs. (6)--(12) it follows that the algebra of these operators has
the form
\begin{eqnarray}
 &&\bar{\Delta}^{\{a}\bar{\Delta}^{b\}}=0,\\
 &&\tilde{\Delta}^{\{a}\tilde{\Delta}^{b\}}=0,\\
 &&\bar{\Delta}^{\{a}\tilde{\Delta}^{b\}}+
 \tilde{\Delta}^{\{a}\bar{\Delta}^{b\}}=0.
\end{eqnarray}

\section{Generating Functional, Extended BRST \\Symmetry and Gauge
Independence}

Let us denote by $S=S(\phi,\phi^*,\pi,\bar{\phi})$ the quantum action,
corresponding to the initial classical theory with the action $S_0$, and
defined as a solution of the following master equations:
\begin{eqnarray}
 \frac{1}{2}(S,S)^a+V^aS=i\hbar\Delta^aS,
\end{eqnarray}
with the standard boundary condition
\begin{eqnarray}
 \left.S\right|_{\phi^*=\bar{\phi}=\hbar=0}=S_0.
\end{eqnarray}
Eq.~(20) can be represented in the equivalent form
\begin{eqnarray}
 \bar{\Delta}^a\exp\left\{\frac{i}{\hbar}S\right\}=0.
\end{eqnarray}
Let us further define the vacuum functional as the following functional
integral:
\begin{eqnarray}
 Z_X=\int d\phi\,d\phi^*d\pi\,d\bar{\phi}\,d\lambda\exp\left\{
 \frac{i}{\hbar}\left(S+X+\phi^*_{Aa}\pi^{Aa}\right)\right\},
\end{eqnarray}
where $X=X(\phi,\phi^*,\pi,\bar{\phi},\lambda)$ is a bosonic functional
depending also on the new variables $\lambda^A$,
$\varepsilon(\lambda)=\varepsilon_A$, which serve as gauge-fixing
parameters. We require that the functional $X$ satisfies
 the following master equation:
\begin{eqnarray}
 \frac{1}{2}(X,X)^a-U^aX=i\hbar\Delta^aX,
\end{eqnarray}
or, equivalently,
\begin{eqnarray}
 \tilde{\Delta}^a\exp\left\{\frac{i}{\hbar}X\right\}=0.
\end{eqnarray}
Notice that the generating equations determining the quantum action $S$ in
eq.~(20) (or (22)) and the gauge-fixing functional $X$ in eq.~(24)
(or (25)) differ---along with the vacuum functional $Z$ in
eq.~(23)---from the corresponding definitions applied in the method of
TQ \cite{12a}.

One can easily obtain the simplest solution of eq.~(24) (or eq.~(25))
determining the gauge-fixing functional $X$
\begin{eqnarray}
 X&=&\left(\bar{\phi}_A-\frac{\delta F}{\delta\phi^A}\right)\lambda^A-
 \frac{1}{2}\varepsilon_{ab}U^aU^bF=\nonumber\\
 &=&\left(\bar{\phi}_A-\frac{\delta F}{\delta\phi^A}\right)\lambda^A
 -\frac{1}{2}\varepsilon_{ab}\pi^{Aa}{\frac {\delta^2 F}{\delta\phi^A
 \delta\phi^B}}\pi^{Bb},
\end{eqnarray}
where $F=F(\phi)$ is a bosonic functional depending only on the fields
$\phi^A$. As a straightforward exercise, one makes sure that the
functional $X$ in eq.~(26) does satisfy eq.~(24). If we further demand
that the quantum action $S$ does not depend on the fields $\pi^A$, then
the functional (26) becomes exactly the vacuum functional of the $Sp(2)$
quantization scheme \cite{7a,8a}.

Let us consider a number of properties inherent in the present scheme of
triplectic quantization, i.e. modified according to eq.~(20)--(25).
In the first place, the vacuum functional (23) is invariant under the
following transformations:
\begin{eqnarray}
 \delta\Gamma=(\Gamma,-S+X)^a\mu_a+\mu_a(V^a+U^a)\Gamma,
\end{eqnarray}
where $\mu_a$ is an $Sp(2)$ doublet of constant anticommuting parameters,
and $\Gamma$ stands for any of the variables $\phi$, $\phi^*$, $\pi$,
$\bar{\phi}$. Eq.~(27) defines the
transformations of extended BRST symmetry, realized on the space of the
variables $\phi$, $\phi^*$, $\pi$, $\bar{\phi}$. In the particular case,
corresponding to the gauge-fixing boson chosen as in eq.~(26), we have
\begin{eqnarray}
 \delta\phi^A&=&-\left(\frac{\delta S}{\delta\phi^*_{Aa}}
 -\pi^{Aa}\right)\mu_a,\\
 \delta\phi^*_{Aa}&=&\mu_a\left(\frac{\delta S}{\delta\phi^A}
 +\frac{\delta^2F}{\delta\phi^A\delta\phi^B}\lambda^B
 +\frac{1}{2}(-1)^{\varepsilon_A}\varepsilon_{bc}\pi^{Bb}
 \frac{\delta^3F}{\delta\phi^B\delta\phi^A\delta\phi^C}\pi^{Cc}\right)\!,\\
 \delta\pi^{Aa}&=&\varepsilon^{ab}\left(
 \frac{\delta S}{\delta\bar{\phi}_A}
 -\lambda^A\right)\mu_b,\\
 \delta\bar{\phi}_A&=&\mu_a\varepsilon^{ab}\left(
 \frac{\delta S}{\delta\pi^{Ab}}+\phi^*_{Ab}\right)
 +\mu_a\frac{\delta^2F}{\delta\phi^A\delta\phi^B}\pi^{Ba}\,.
\end{eqnarray}

Consider now the question of gauge dependence in the case of the vacuum
functional $Z$, eq.~(23). Any admissible variation $\delta X$ should satisfy
the equations
\begin{eqnarray}
 (X,\delta X)^a-U^a\delta X=i\hbar\Delta^a\delta X.
\end{eqnarray}
It is convenient to consider an $Sp(2)$ doublet of operators
$\hat{S}^a(X)$, defined by the rule
\begin{eqnarray}
 (X,F)^a\equiv\hat{S}^a(X)\cdot F,
\end{eqnarray}
and possessing the properties
\begin{eqnarray}
 \hat{S}^{\{a}(X)\hat{S}^{b\}}(X)=\hat{S}^{\{a}\left
 (\frac{1}{2}(X,X)^{b\}}\right),
\end{eqnarray}
which follow from the generalized Jacobi identities (2). Eq.~(32) can be,
consequently, represented in the form
\begin{eqnarray}
 \hat{Q}^a(X)\delta X=0,
\end{eqnarray}
where we have introduced an $Sp(2)$ doublet of operators $\hat{Q}^a$,
defined by the rule
\begin{eqnarray}
 \hat{Q}^a(X)=\hat{S}^a(X)-i\hbar\tilde{\Delta}^a.
\end{eqnarray}
With allowance for eq.~(24) the operators $\hat{Q}^a$ form a set of
nilpotent anticommuting operators, i.e.
\begin{eqnarray}
 \hat{Q}^{\{a}(X)\hat{Q}^{b\}}(X)=0.
\end{eqnarray}
By virtue of eq.~(37), any bosonic functional of the form
\begin{eqnarray}
 \delta X=\frac{1}{2}\varepsilon_{ab}\hat{Q}^a(X)\hat{Q}^b(X)\delta Y,
\end{eqnarray}
with an arbitrary bosonic functional $\delta Y$, is a solution of eq.~(35).
Moreover, by analogy with the theorems proved in ref.~\cite{9a}, one
establishes the fact that any solution of eq.~(35)---vanishing when all the vari
in $\delta X$ are equal to zero---has the form (38), with a certain
bosonic functional $\delta Y$. In the particular case of the gauge
functional $X$ (26), its variation $\delta X$ can be easily represented
in the form of eq.~(38), i.e.
\begin{eqnarray}
 \delta X=-\frac{\delta(\delta F)}{\delta\phi^A}\lambda^A
 -\frac{1}{2}\varepsilon_{ab}\pi^{Aa}{\frac{\delta^2(\delta F)}
 {\delta\phi^A\delta\phi^B}}\pi^{Bb}
 =-\frac{1}{2}\varepsilon_{ab}\hat{Q}^a(X)\hat{Q}^b(X)\delta F
\end{eqnarray}
with $\delta Y=-\delta F$.

Let us denote by $Z_X\equiv Z$ the value of the vacuum functional (23)
corresponding to the gauge condition chosen as a functional $X$.

In the vacuum functional $Z_{X+\delta X}$ we first make the change of
variables (27), with $\mu_a=\mu_a(\Gamma,\lambda)$, and then,
accompanying it with a subsequent change of variables
\begin{eqnarray}
 \delta\Gamma=(\Gamma,\delta Y_a)^a,\;\;\;\varepsilon(\delta Y_a)=1,
\end{eqnarray}
with $\delta Y_a=-i\hbar\mu_a(\Gamma,\lambda)$, we arrive at
\begin{eqnarray}
 Z_{X+\delta X}=\int d\phi\,d\phi^*d\pi\,d\bar{\phi}\,d\lambda
 \exp\left\{\frac{i}{\hbar}\bigg(S+X+\delta X+\delta X_1
 +\phi^*_{Aa}\pi^{Aa}\bigg)\right\},
\end{eqnarray}
In eq.~(41) we have used the notation
\begin{eqnarray}
 \delta X_1=2\bigg((X,\delta Y_a)^a-U^a\delta Y_a-i\hbar\Delta^a\delta
 Y_a\bigg)=2\hat{Q}^a(X)\delta Y_a.
\end{eqnarray}
Let us choose the functional $\delta Y_a$ in the form
\begin{eqnarray}
 \delta Y_a=\frac{1}{4}\varepsilon_{ab}\hat{Q}^b\overline{\delta Y},\;\;\;
 \varepsilon(\overline{\delta Y})=0.
\end{eqnarray}
Then, representing $\delta X$ as in eq.~(38), and identifying
$\delta Y=-\overline{\delta Y}$, we find that
\begin{eqnarray}
 Z_{X+\delta X}=Z_X,
\end{eqnarray}
i.e. the vacuum functional (and hence, by virtue of the equivalence
theorem \cite{14a}, also the $S$ matrix) does not depend on the choice
of gauge. Note that in the particular case (39) we have
$\overline{\delta Y}=\delta F$.

\section{Concluding Remarks}
The reader may profit by considering the original \cite{12a} version
of TQ as compared to the modified scheme,
proposed in the this paper. Thus, both versions are based on extended
BRST symmetry. Both versions apply the vacuum functional and the $S$
matrix not depending on the choice of gauge, while admitting of gauges
which reproduce the results of the $Sp(2)$ symmetric quantization. Both
versions implement the idea of separate treatment of the quantum action
and the gauge-fixing functional, based each on appropriate master
equations. The principal distinctions concern a different form
of these equations as well as a different form of the vacuum
functional. The modification of the generating equations \cite{12a} permits
incorporating the information contained in the initial classical action by
means of the corresponding boundary conditions. In contrast to the original
version \cite{12a}, the classical action provides a solution of the modified
master equation. Thus, one establishes a connection with the previous
schemes of LQ. In particular, one easily reveals the
fact of equivalence with the $Sp(2)$ quantization, by means of explicit
realization of the corresponding class of boundary condition. In the original
version of TQ, however, these questions still remain open. Another
distinction of the two TQ schemes is connected with the explicit structure of
the corresponding master equations. Thus, the original version \cite{12a} of
TQ defined the generating equations for the quantum action and the vacuum
functional, using the operators
\begin{eqnarray}
 V_{\rm BM}^a=\frac{1}{2}\left(
 \varepsilon^{ab}\phi^{*}_{Ab}\frac{\delta}{\delta\bar\phi_A}
 +\pi^{Aa}(-1)^{\varepsilon_A+1}\frac{\delta_l}{\delta\phi^A}\right)
 =\frac{1}{2}(U^a+V^a).
\end{eqnarray}
The use of the generating equations determining the quantum action with
the help of the operators $V_{\rm BM}^a$ leads to the following
characteristic feature of the triplectic quantization \cite{12a}: the
classical action of the initial theory, defined as a limit of the quantum
action at $\hbar\to 0$ and $\phi^*=\bar{\phi}=\pi=0$, does not satisfy the
generating equations of the method. In turn, the proofs of the
existence theorems for the generating equations in all known methods
of LQ are based on the fact that the initial
classical action is a solution of the corresponding master equations.
Moreover, form the viewpoint of the superfield quantization \cite{13a},
which applies operators $V^a$, $U^a$, whose component representation is
\begin{eqnarray}
 V^a&=&\varepsilon^{ab}\phi^{*}_{Ab}\frac{\delta}{\delta\bar\phi_A}
 -J_A\frac{\delta}{\delta\phi^{*}_{Aa}},\nonumber\\
 \\
 U^a&=&(-1)^{\varepsilon_A+1}\pi^{Aa}
 \frac{\delta_l}{\delta\phi^A}
 + (-1)^{\varepsilon_A}\varepsilon^{ab}\lambda^A
 \frac{\delta_l}{\delta\pi^{Ab}}\nonumber
\end{eqnarray}
(with $J_A$ being the sources to the fields $\phi^A$), the operators (45)
have no precise geometrical meaning, whereas the $V^a$ and $U^a$ in
eq.~(46) serve as generators of supertranslations---in superspace spanned
by superfields and superantifields---along additional (Grassmann)
coordinates. In turn, the operators $V^a$ (4) and $U^a$ (5), applied in
this paper, can be considered as limits (at $J_A=0$, $\lambda^A=0$) of the
operators (46), which possess a clear geometrical meaning. The present
modified scheme of triplectic quantization enjoys every attractive feature
of the quantization \cite{12a}: the theory possesses extended BRST
transformations; the vacuum functional and the $S$ matrix do not depend on
the choice of the gauge-fixing functional; there exists such a choice
of the gauge-fixing functional and solutions of the generating
equations that reproduces the results of the $Sp(2)$ method.

\section*{Acknowledgments}
The work of one of the authors (PML) has been supported by the Russian
Foundation for Basic Research (RFBR), project 96--02--16017, as well as
by grant INTAS 96-0308 and by grant RFBR-DFG 96--02--00180. PML
appreciates the kind hospitality extended to him by the Center of
Advanced Study (NTZ) of Leipzig University and by the Institute
of Physics of the University of S\~{a}o Paulo. DMG thanks
Brazilian Foundation CNPq and FAPESP for partial support.


\end{document}